\documentclass[letterpaper,twocolumn]{jpsj3}
\usepackage{txfonts}

\usepackage{siunitx}
\usepackage{textcomp}

\usepackage{algpseudocode}
\usepackage{algorithm}

\usepackage{here}

\usepackage{tabularx}

\usepackage{graphicx}
\usepackage{float}
\usepackage{textcomp}
\usepackage{amsmath}
\usepackage{listings}
\usepackage{url}

\usepackage{tikz}
\usetikzlibrary{trees}

\title{Reconsideration of optimization for reduction of traffic congestion}

\author{Masayuki Ohzeki}
\inst{Graduate School of Information Sciences, Tohoku University, Miyagi 980-8564, Japan \\
Department of Physics, Tokyo Institute of Technology, Tokyo, 152-8551, Japan \\
Sigma-i Co., Ltd., Tokyo, 108-0075, Japan
} 

\abst{
One of the most impressive applications of a quantum annealer was optimizing a group of Volkswagen to reduce traffic congestion using a D-Wave system.
A simple formulation of a quadratic term was proposed to reduce traffic congestion.
This quadratic term was useful for determining the shortest routes among several candidates.
The original formulation produced decreases in the total lengths of car tours and traffic congestion.
In this study, we reformulated the cost function with the sole focus on reducing traffic congestion.
We then found a unique cost function for expressing the quadratic function with a dead zone and an inequality constraint.
}

\begin{document}
\maketitle
\renewcommand{\algorithmicrequire}{\textbf{Input:}}
\renewcommand{\algorithmicensure}{\textbf{Output:}}

Quantum annealing (QA) can be used to solve a combinatorial optimization problem \cite{kadowaki1998}.
The physical implementation of the transverse Ising model enabled us to perform QA experiments.
Since the appearance of quantum annealers, many applications have exhibited their potential effects in various fields such as traffic flow optimization \cite{neukart2017traffic},
finance \cite{rosenberg2016solving}, logistics \cite{Ohzeki2019control, mugel_dynamic_2022}, manufacturing \cite{venturelli2016quantum, Haba2022}, material pre-processing \cite{Tanaka2023}, the search for nontrivial candidates \cite{Doi2023}, marketing \cite{Nishimura2019}, maze generation \cite{Ishikawa2023}, and decoding problems \cite{Ide2020, Arai2021code}.
Fast solvers have also been proposed by integrating them with classical methods for operation research, as described in the literature \cite{Hirama2023, takabayashi2024}.
Model-based Bayesian optimization using quantum annealing has also been proposed \cite{Koshikawa2021, morita2023}.
A comparative QA study was performed using benchmark tests to solve optimization problems \cite{Oshiyama2022}. 
Because environmental effects cannot be avoided, quantum annealers are sometimes considered simulators of quantum many-body dynamics \cite{Bando2020, Bando2021, King2022}. 
Quantum annealers are also available in machine learning, particularly for Boltzmann machines \cite{ Amin2018, Arai2021, Sato2021, Urushibata2022, Goto2023, hasegawa2023}.

This study introduces an interesting formulation of the quadratic unconstrained binary optimization (QUBO) problem, a special input form for quantum annealers.
The penalty method is typically used to formulate the QUBO problem to satisfy the equality constraints.
The $K$-hot penalty method is often used in cost functions (\ref{ori}) and (\ref{ren}).
However, formulating a QUBO problem to satisfy the inequality constraints is not a straightforward process \cite{Yonaga2022}.
Redundant binary variables are often employed to represent the inequality constraints.
Because the number of qubits tends to be large, we avoid implementing inequality constraints in formulating a QUBO problem suitable for quantum annealers, which have a limited number of qubits.
However, the ``half-hot" penalty method realizes the inequality constraints.
Let us consider the following optimization problem:
\begin{equation}
    \min_{\bf x}\left\{ f({\bf x}) \right\} {\rm s.t.} \sum_{i=1}^N \sum_{\mu \in M_i} C_{e\mu i}x_{\mu i} \le 1\quad e \in E.
\end{equation}
We can rephrase this optimization problem using the ``half-hot" penalty method:
\begin{equation}
    \min_{\bf x}\left\{ f({\bf x}) + \frac{\lambda}{2}\sum_{e \in E}\left( \sum_{i=1}^N \sum_{\mu \in M_i} C_{e\mu i}x_{\mu i} - \frac{1}{2} \right)^2 \right\}.
\end{equation}
This is because $y=0$ and $y=1$ yield the same value, $(y-1/2)^2$.
The first proposal for formulating QUBO using the ``half-hot" penalty method in this context can be found in the literature \cite{Haba2022}.
The resulting penalty term is a quadratic function with a dead zone.
We use only the quadratic function in the formulation of the QUBO problem.
Thus, the variety of models expressed in the QUBO form is limited.
However, the "half-hot" penalty method allows a little-bit wide range of optimization problems to be modeled in the QUBO form.
In addition, we found a fault in the formulation of a well-known and celebrated application of quantum annealers.

The most significant application of quantum annealers is traffic congestion reduction \cite{neukart2017traffic}.
A quadratic term that squares the number of cars passing through the road segments was proposed.
In short, we reviewed the reduction in traffic congestion.
We define binary variables $x_{\mu i} \in {0,1}$ for each car $i$.
The label $\mu \in M_i$ denotes the index of the routes emanating from the starting point to the destination of car $i$, where $M_i$ is a set of candidate routes for car $i$.
In addition, we set $C_{e\mu i} \in {0,1}$ for each road segment $e$ and route $\mu$ for car $i$.
Then, the cost function is defined as follows:
\begin{equation}
    E({\bf x}) = \sum_{e \in E} \left( \sum_{i=1}^N \sum_{\mu \in M_i} C_{e\mu i}x_{\mu i} \right)^2 + \frac{\lambda}{2} \sum_{i=1}^N \left( \sum_{\mu \in M_i} x_{\mu i} -1 \right)^2,\label{ori}
\end{equation}
where $\lambda$ is the coefficient of the penalty term satisfying the equality constraint.
Here, the summation term, $\sum_{i=1}^N \sum_{\mu \in M_i} C_{e\mu i}x_{\mu i} \equiv C_e $, denotes the number of cars passing through road segment $e$.
If no car passes through segment $e$, $C_e = 0$.

Why is a quadratic term used in cost functions?
The quadratic term indicates an increase in $C_e$.
However, in the strict sense, only a single car passes through road segment $C_e=1$, which affects the cost function.
Thus, if we minimize the original cost function, the optimal solution does not allow a long tour given by a collection of road segments, $e$, through which only a single car travels.
In other words, the total tour length should be minimized by optimizing the original cost function.
Therefore, the original formulation reduces traffic congestion and minimizes the total lengths of the tours for all the cars.

This was again found by considering the following computation.
We extract half from $C_e$ in the quadratic term and expand it as follows:
\begin{equation}
    \sum_{e \in E} C_e^2 = \sum_{e \in E} \left( C_e - \frac{1}{2} \right)^2 + \sum_e C_e + \frac{1}{4}. \label{ren}
\end{equation}
The first term on the right-hand side is a correct formulation for reducing traffic congestion. It is the same as the ``half-hot" penalty method.
This yields no difference between cases $C_e=0$ and $C_e=1$.
The first term is a quadratic function with a dead zone.
The case where $C_e=1$ indicates that only a single car passes through road segment $e$.
However, this has not resulted in traffic congestion.
The case in which $C_e > 1$ increases the possibility of traffic congestion on road segment $e$.
Thus, minimizing the first term on the right-hand side of Eq. ~ (\ref{ren}) reduces traffic congestion.
The second term on the right side of Eq. (\ref{ren}) is the same as the total distance between all the cars.
Its minimization decreases the total distance and encourages the shortest route for each car.

After collecting all the above recommendations, the corrected cost function for reducing traffic congestion is as follows:
\begin{equation}
    E({\bf x}) = \sum_{e \in E} \left( \sum_{i=1}^N \sum_{\mu \in M_i} C_{e\mu i}x_{\mu i} - \frac{1}{2} \right)^2 + \frac{\lambda}{2} \sum_{i=1}^N \left( \sum_{\mu \in M_i} x_{\mu i} -1 \right)^2.\label{after}
\end{equation}
We find that Eq. (\ref{after}), rather than Eq. (\ref{ori}), is better at reducing traffic congestion if our formulation is applied to real car traffic map data.
If the goal is to shorten the distances of all the car tours, a linear term can be added to the cost function (\ref{ren}) with a small value for coefficient $\alpha$.
The original formulation is essentially identical to $\alpha=1$.

The formulation of the QUBO problem does not necessarily have rich descriptive power because the model is quadratic.
In particular, it is difficult to realize inequality constraints in the QUBO form.
In addition, only the simple shape of the quadratic function is realized.
However, as in this study, a little ingenuity opens up the applicability of the QUBO form.

{\it Acknowledgments }. 
This study was financially supported by programs for bridging the gap between R\&D and IDeal society (Society 5.0) and Generating Economic and social value (BRIDGE) and Cross-ministerial Strategic Innovation Promotion Program (SIP) from the Cabinet Office.

\bibliographystyle{jpsj}
\bibliography{jpsj_template}
\end{document}